\documentclass{svjour3}                     
\smartqed  
\usepackage{graphicx,aas_macro_long,natbib}
\begin{document}
\title{Solar irradiance models and measurements: a comparison in the 220~nm to 240~nm 
wavelength band}

\titlerunning{Solar irradiance models and measurements} 

\author{Yvonne C. Unruh \and
        Will T. Ball \and Natalie A. Krivova
}


\institute{Y. C. Unruh \at
              Astrophysics Group, Blackett Laboratory, Imperial College London, SW7 2AZ, UK \\
              \email{y.unruh@imperial.ac.uk}           
           \and
           W. Ball \at
              Astrophysics Group, Blackett Laboratory, Imperial College London, SW7 2AZ, UK \\
              \email{william.ball08@imperial.ac.uk}           
           \and
	   N. A. Krivova \at
	      Max-Planck Institut f\"ur Sonnensystemforschung, D-37191 Katlenburg-Lindau, Germany \\
	      \email{natalie@mps.mpg.de}
}

\date{Received: 25$^{\rm th}$ July 2011 / Accepted: date}

\maketitle

\begin{abstract}
Solar irradiance models that assume solar irradiance variations to be due to 
changes in the solar surface magnetic flux have been successfully used to 
reconstruct total solar irradiance on rotational as well as cyclical and 
secular time scales. 
Modelling spectral solar irradiance is not yet as advanced, and also suffers 
from a lack of comparison data, in particular on solar-cycle time scales. 
Here we compare solar irradiance in the 220~nm to 240~nm band as modelled 
with SATIRE-S and measured by different instruments on the UARS and 
SORCE satellites. 

We find good agreement between the model and measurements on rotational 
time scales. The long-term trends, however, show significant differences. 
Both SORCE instruments, in particular, show a much steeper gradient over 
the decaying part of cycle 23 than the modelled irradiance or that measured 
by UARS/SUSIM.
\end{abstract}

\section{Introduction}
\label{intro}
The solar irradiance, i.e., the solar energy flux received at the top of 
the Earth's atmosphere, 
changes on a wide range of time scales. Changes in total solar irradiance 
(TSI) have been monitored since 1978 and a number of composites combining data 
from different satellites are available (see, e.g., \cite{domingo2009issi} for a discussion
of TSI composites). Detailed modelling of the TSI has shown that solar irradiance changes
on rotational to cyclical time scales can be well modelled as being 
due to changes in the solar surface magnetic flux alone 
\citep[e.g.][]{preminger-et-al-2002,wenzler2006,krivova2003_cycle23,ball2011}. 

Quantifying changes in solar spectral irradiance (SSI) is a more arduous task. 
Up until the launches of the SCIAMACHY imaging spectrometer onboard ENVISAT 
and the SOlar Radiation and Climate Experiment (SORCE) satellite in 2002 
and 2003, respectively, most of the spectral data were restricted to 
wavelengths below 400~nm, and more reliable data were only available for 
wavelengths below 300~nm \citep{woods-et-al-96}. A compilation of UV 
irradiance measurements up to 2005, though not yet including SORCE data, was 
presented by \cite{deland2008}. This compilation includes the authors' best 
estimate of which of the instruments yield the most reliable measurements 
at different times over the last three solar cycles, though no 
wavelength-by-wavelength cross-calibrations are applied at this stage.

The SOLar STellar Irradiance Comparison Experiment (SOLSTICE)
\citep{mcclintock2005design,mcclintock2005calib} and Spectral 
Irradiance Monitor (SIM) \citep{harder2005design,harder2005calib} 
onboard SORCE measure spectral solar irradiance 
between 115~nm and 320~nm and between 200~nm and 2.4~$\mu$m, respectively. 
Results from the decaying phase of cycle~23 have shown 
rather unexpected long-term behaviour, in particular a much steeper UV 
decline than that measured in cycle 22 by the Solar Ultraviolet Spectral 
Irradiance Monitor (SUSIM) and by SOLSTICE onboard UARS (Upper Atmosphere 
Research Satellite). In addition, SORCE/SIM also suggested an anticorrelation 
of SSI between wavelengths of $\sim 400$~nm to $\sim 700$~nm with TSI \citep{harder2009}. 

Model-based reconstructions of the solar UV irradiance have been presented by a number 
of authors, including \cite{krivova2009uv1974,krivova2006UV} and \cite{morrill2011}. 
Here we present reconstructions of the UV irradiance over the last 
solar cycle for wavelengths between 220~nm and 240~nm using the SATIRE-S 
(Spectral And Total Irradiance REconstructions - Satellite era) model. 
We compare the UV reconstructions with data from UARS/SUSIM and 
UARS/SOLSTICE, as well as to data from SIM and SOLSTICE onboard SORCE. 
A more in-depth analysis covering a wider spectral range as well as solar cycles 21 and 
22 will be presented in Ball et al (in prep). 

\section{SATIRE-S} 
\label{sec:satire}
The reconstructions presented here are obtained with SATIRE-S \citep{krivova2011review}
and assume that all changes in solar irradiance are due to changes in the solar surface 
magnetic flux. The emergent magnetic flux forms dark and bright surface features that 
can be identified from magnetograms and continuum images. Having identified the magnetic 
features and knowing their contrast with respect to the quiet Sun \citep[see][]{unruh99lumi}, 
the irradiance change can be calculated by summing the different contributions over the solar surface. 

SATIRE-S has one free parameter, $B_{\rm sat}$; it effectively determines the field
strength above which the facular contrast saturates and depends on the resolution 
and image characteristics of the magnetograms and continuum images (see 
\citeauthor{krivova2011review}, \citeyear{krivova2011review} for more detail). 

\section{Solar UV irradiance from 1999 to 2010}
The reconstructions presented here are based on magnetograms and continuum 
images from the Michelson Doppler Imager (MDI) onboard the Solar and 
Heliospheric Observatory (SoHO). We consider the period between 1999 and 
2009 after the SoHO `vacation'. During this time we have previously found 
excellent agreement between 
the PMOD TSI composite \citep{frohlich2006pmod_revision21,frohlich2000SSR} and the SATIRE-S 
reconstructions. The free parameter, $B_{\rm sat}$, was fixed so that the regression slope of 
the TSI values calculated with SATIRE-S and those of the PMOD composite was unity for times 
between February 1999 and February 2010. The resulting value for $B_{\rm sat}$ is 424~G and 
the correlation coefficient over this period is $r = 0.983$ ($r^2 = 0.966$). Note that this 
value of $B_{\rm sat}$ can not be compared directly to earlier values 
\citep[e.g.][]{krivova2003_cycle23} on account of the changes in the magnetogram calibrations 
introduced for the MDI level 1.8~data. 

\begin{figure}
	\includegraphics[width=0.92\textwidth]{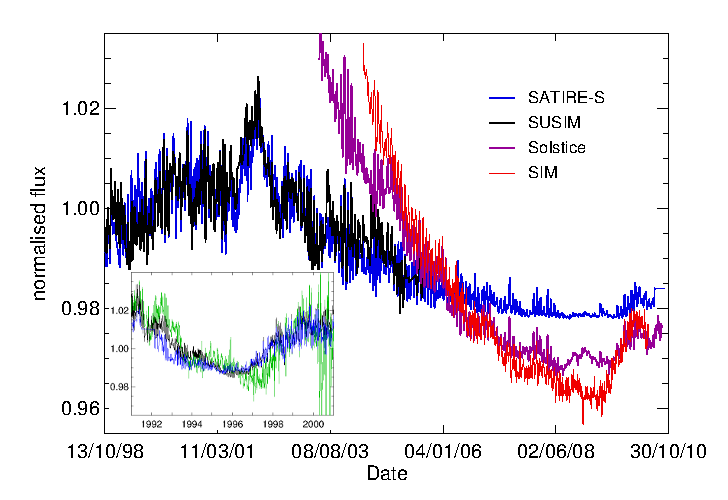}
        \caption[]{Plot of the normalised solar UV irradiance integrated between 220~nm
        and 240~nm as modelled with SATIRE-S (blue line) and measured with UARS/SUSIM (black),
        SORCE/SOLSTICE (purple) and SORCE/SIM (red). The inset shows data going back to 
	1991: SATIRE-S (blue), UARS/SUSIM (black) and UARS/SOLSTICE (green).}
        \label{fig:uv_long}       
\end{figure}

The UV irradiances integrated between 220~nm and 240~nm are shown in Fig.~\ref{fig:uv_long}
for times between 1999 and 2010. This spectral region has been selected as it had previously 
been  found that the SATIRE-S calculations reproduce the UARS/SUSIM measurements well, despite
SATIRE's reliance on LTE which tends to be a poor approximation in this 
wavelength region. For other UV wavelengths, irradiance changes can be derived 
from the integrated 220~nm to 240~nm irradiance by linear regression \citep{krivova2006UV}. 

The SATIRE-S reconstructions are shown in blue, the UARS/SUSIM measurements are plotted in black. The error in the absolute measurements is significantly 
larger than the error in the relative changes of the irradiance 
\citep[see, e.g.,][]{woods-et-al-96}. 
The SUSIM and SATIRE-S data have thus been normalised so that their mean 
values after the SoHO vacation agree. The SORCE/SIM 
and SORCE/SOLSTICE data are overplotted in red and purple, respectively. 
They have been normalised using the mean values during their overlap 
period (May 2005 to May 2010); they have also been shifted by 0.02 for ease of 
comparison with SUSIM and SATIRE. The inset shows a comparison between 
UARS/SOLSTICE (green), UARS/SUSIM (black) and SATIRE-S (blue, based on Kitt
Peak images pre-1999) during the lifetime of the UARS/SOLSTICE instrument. 
As the data quality of UARS/SOLSTICE deteriorated from about 
2001 onwards (as, e.g., indicated by the increase in spurious downward spikes), 
the irradiances shown on the inset have been normalised to their 
pre-2000 mean.

The UV irradiance measured with both SORCE instruments shows a much steeper 
long-term trend than that found with SUSIM, either during cycle 23 or 
cycle 22 (see \cite{krivova2009uv1974} for an in-depth discussion of UV irradiance 
measurements during cycle 22). 

\begin{figure}
        \includegraphics[width=0.9\textwidth,viewport= 50 370 570 710]{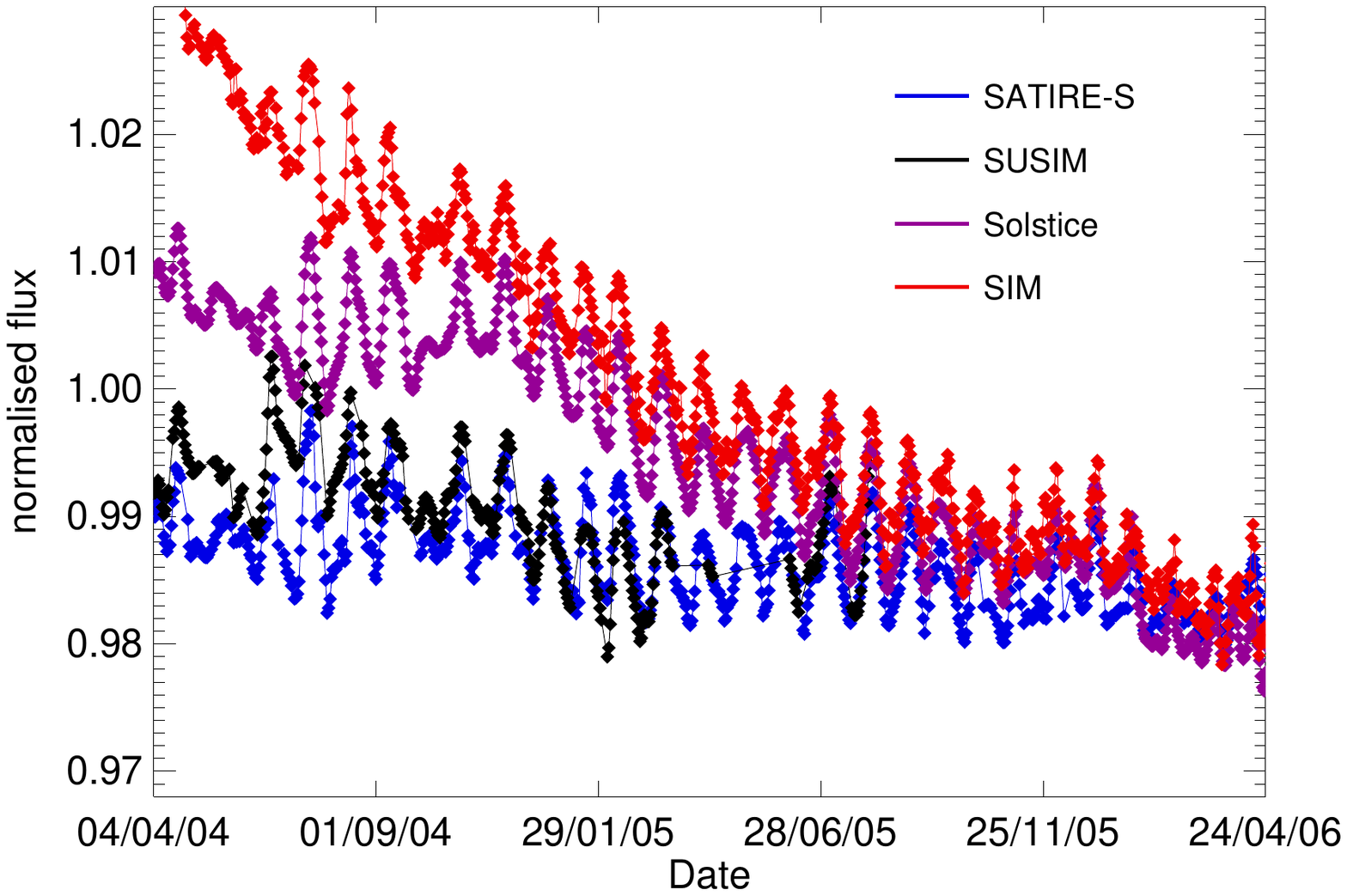}
        \caption[]{A two-year section of the normalised solar UV irradiance. 
	The quantities plotted are as in Fig.~1, but showing the period between 
	April 2004 and April 2006 when UARS and 
	SORCE measurements overlap in more detail.}
        \label{fig:uv}       
\end{figure}

\section{Discussion}
\subsection{Rotational time scales}
The modelled and the measured UV irradiances show excellent agreement on rotational 
time scales. This is illustrated in more detail in Fig.~\ref{fig:uv}, where the 
irradiances are plotted over the shorter timespan of approximately two years. 
The irradiance changes by approximately 1\% as the active regions 
rotate in and out of view and we find that the variability amplitude is well matched 
between the different instruments and the model. Indeed, on rotational time scales, the 
scatter between the irradiances measured with the different instruments is comparable to 
the scatter between the modelled and observed irradiance, as illustrated by the results 
of a regression analysis. The slopes and correlation coefficients are listed in 
Tab.~\ref{tab:rotational}. Rows 2 to 7 are for 
detrended data sets where the cyclical variability has been removed and only rotational 
variability retained. The number of data points for which the 
correlation was calculated is also listed in Tab.~\ref{tab:rotational}; note 
that the UARS/SUSIM measurements cover a much longer timespan compared to SORCE/SIM. 
UARS/SOLSTICE data have not been included in the analysis here as they do not 
overlap with SORCE/SIM. We note, however, that the UARS/SUSIM and UARS/SOLSTICE 
rotational variability agree well with each other and also with SATIRE-S.

\begin{table}
	\caption{Correlation coefficients (column 4), regression slopes (column 5) and number of 
	overlapping data points considered (column 6) for the SATIRE-S, UARS/SUSIM, 
	SORCE/SIM and SORCE/SOLSTICE data. The results in the first row are for the normalised
	data and include the long-term cyclical trend; the remaining rows are all for detrended 
	data. 
	}
	\label{tab:rotational}    
	\begin{tabular}{cllllrc}
	   \hline\noalign{\smallskip}
	   [1]	& \ [2]	& \ [3]		& \ [4]		& \ [5]	& [6] \	& [7] \\
	   row	& data set 1 & data set 2 & $r$ ($r^2$) & slope & points & Time span \\
	   \noalign{\smallskip}\hline\noalign{\smallskip}
	   1	& SATIRE-S 	& SUSIM	&  0.92 (0.85)	& 1.03 	& 1641 & 1999/02/19 -- 2005/07/31 \\
	   \noalign{\smallskip}\hline\noalign{\smallskip}
	   2	& SATIRE-S 	& SUSIM 	& 0.88 (0.77)   & 1.04      & 1641 & 1999/02/19 -- 2005/07/31 \\
	   3	& SATIRE-S	& SIM		& 0.76 (0.58)	& 1.14	& 1665 & 2004/04/22 -- 2009/10/30 \\
	   4	& SATIRE-S	& SOLSTICE 	& 0.91 (0.83)	& 0.97	& 523 & 2003/05/14 -- 2005/07/31 \\
	   5	& SUSIM		& SOLSTICE	& 0.75 (0.56)	& 0.89	& 523 &	2003/05/14 -- 2005/07/31 \\
	   6	& SUSIM		& SIM		& 0.72 (0.52)	& 0.96	& 293 & 2004/04/22 -- 2005/07/31 \\
	   7	& SIM		& SOLSTICE	& 0.80 (0.64)	& 0.91	& 1665 & 2004/04/22 -- 2009/10/30 \\ 
	   \noalign{\smallskip}\hline
	\end{tabular}
\end{table}

On rotational timescales, we find that the best agreement is between SATIRE-S and 
SORCE/SOLSTICE; the model is able to explain at least 83\% of the rotational 
variability.  This suggests that, on these timescales and for the instruments 
considered here, SORCE/SOLSTICE has the lowest instrumental noise. The higher 
instrumental noise of SIM is not too surprising as the $220 - 240$~nm wavelength 
band is relatively near the blue edge of the SIM instrument 
\citep[see also the discussion in][]{unruh2008}. 
Considering the instrument designs and capabilities, SORCE/SOLSTICE should 
yield better results for wavelengths below 240~nm, while SORCE/SIM is 
the instrument of choice above approximately 290~nm (J Harder, priv comm). 
Probably due to the loss of one of the reaction wheels in 2008, 
the noise level of SORCE/SIM increased slightly towards the end of the mission
(see the period around 
cycle minimum in Fig.~\ref{fig:uv_long}). This is borne out by the 
correlation analysis. Restricting the analysis to times between April 
2004 and January 2008 results in higher correlation coefficients 
($r = 0.81$ vs $0.76$ for the comparison with SIM and 
$r=0.84$ vs $0.80$ for the comparisons with and SOLSTICE). 
At the same time, the correlation slopes improve to values nearer unity; in 
the case of SATIRE-S vs SIM (see row 3) the slope decreases from 
1.14 to 1.05, while it increases from 0.91 to 0.97 for the SIM to 
SOLSTICE comparison (see row 7). 

The rather low correlation coefficient for UARS/SUSIM with either 
SORCE/SIM or SORCE/SOLSTICE is mainly due to the short time-span 
considered and due to what appears to be increased noise in SUSIM after 
mid 2003. The correlation cofficient between the detrended SATIRE-S 
and SUSIM time series for the time span after 2003 falls to $r=0.79$ 
$(r^2=0.63)$ compared to the more 
typical $r=0.88$ $(r^2=0.77)$ for the period starting in 1999.  

\subsection{Cycle variability}
The most striking feature in Fig.~\ref{fig:uv_long} is the much more pronounced 
long-term trend observed by the two SORCE instruments compared to UARS/SUSIM 
and also to SATIRE-S. As already pointed out by 
\cite{krivova2009uv1974,krivova2006UV} for the descending phase of cycle~22 
and the rising phase of cycle~23, we find that the SATIRE-S reconstructions 
are in excellent agreement with the SUSIM measurements in the 220~nm to 
240~nm range. But while SUSIM and SATIRE-S suggest a difference of the order of 
4\% between cycle minimum and maximum in cycle 22, SORCE/SIM and SOLSTICE show a 
significantly larger change during part of the declining phase of cycle 23. 

The decline observed with SORCE/SOLSTICE is of the order of 6\% between May 2003
and the cycle 23/24 minimum. A similar decline is seen for SORCE/SIM between 
May 2004 and the cycle minimum. We note that in terms of the TSI change, 
the period between May 2003 (and indeed also May 2004) and the cycle minimum 
only corresponds to approximately half of the total cycle change in TSI. 
The $\sim 6$\% decline in the 220 to 240~nm integrated irradiance would thus 
be a lower limit and suggest an overall cycle change of the order of 12\%. This 
is compared to the $\sim 4$\% change in UARS/SUSIM and SATIRE-S (recall that 
the typical rotational variability measured by all instruments out of minimum 
is of the order of 1 to 2\% in this wavelength range).

The relative accuracy of both UARS instruments over a timescale of a few 
years has conservatively been estimated to be of the order of 
1\% \citep{woods-et-al-96}. In the 220~nm to 240~nm region, the
SORCE/SOLSTICE accuracy and long-term stability is considered to be better
than that of SIM \citep{harder-et-al-2010}. \cite{snow-et-al-2005}
estimate the SORCE/SOLSTICE stability to be of the order of 1\% per year
currently, though it is expected that, pending outstanding corrections, 
SORCE/SOLSTICE will achieve a long-term stability of 0.5\% per year by 
the end of the mission (Snow, priv comm).  

We thus find that the cycle variability observed with UARS/SUSIM, 
with UARS/SOLSTICE, and modelled with SATIRE-S agrees within the estimated 
uncertainty. Taking the 1\%-per-year uncertainty at face value, the observed 
trends for SORCE lie just outside the long-term instrumental uncertainties and 
it is not clear how to reconcile the 
UARS and SORCE measurements in this wavelength region.
Note that, while considerably steeper overall, the trends observed with the 
two SORCE instruments do not always agree well. The slope seen in 
SORCE/SIM during the first 9 months is substantially steeper than that 
recorded by SORCE/SOLSTICE. Both instruments show good long and short-term 
agreement over a timespan of approximately two years (February 2005 to 
January 2007) where they both record a decline of approximately 1\% per year. 
Their subsequent trends, however, disagree at the estimated 1\% stability level: 
SORCE/SIM shows a deeper minimum combined with a sharper decline and rise in 
and out of the minimum.

\section{Summary}
The typical rotational variability in the 220~nm to 240~nm band is of the 
order of 1\%. On rotational timescales, we find excellent agreement between 
the irradiances recorded by UARS/SUSIM, SORCE/SOLSTICE and SORCE/SIM, and 
those reconstructed from SoHO/MDI images using the SATIRE-S model. 

On yearly time scales and longer, the observed trends with the newer 
SORCE instruments, SORCE/SOLSTICE and SORCE/SIM, are much more pronounced 
than those measured with UARS/SUSIM or modelled with SATIRE-S. It is 
currently unclear how these trends can be reconciled and whether the differences 
are purely instrumental, or due to a change in the Sun's behaviour. One 
might argue that, after 14 years in space, UARS/SUSIM suffered from 
instrumental degradation during the declining phase of cycle 23. This seems
unlikely, however, when considering the long-term trends measured during the 
second half of cycle 22 that are of a similar magnitude as those in cycle 
23 and, furthermore, agree with those modelled by SATIRE-S. While it is 
simplistic to assume that UV irradiance trends will scale linearly with 
TSI, the approximately fourfold larger drop in the $220-240$~nm irradiance 
during cycle 23 when the change in the TSI was comparable to that in cycle 
22 remains intriguing. 

\begin{acknowledgements}
The authors would like to thank ISSI for the hospitality and lively meeting. 
The authors would also like to thank Jerry Harder for providing the SORCE/SIM 
data, as well as Linton Floyd for helpful discussions and information on SUSIM data 
and Marty Snow for information on the SORCE/SOLSTICE data. 
This work was supported by the
\emph{Deut\-sche For\-schungs\-ge\-mein\-schaft, DFG\/} project
number SO~711/1-3 and by the NERC SolCli consortium grant; we are indebted
to Sami Solanki and Thomas Wenzler for many useful discussions.
\end{acknowledgements}

\bibliographystyle{spbasic}      
\bibliography{unruh_BK}   

\end{document}